\documentclass[aps,pre,twocolumn,superscriptaddress,10pt,floatfix]{revtex4-2}
\usepackage{graphicx}
\usepackage{xcolor}
\usepackage{dcolumn}
\usepackage{amssymb,amsmath,amsfonts,physics}
\usepackage{comment}
\usepackage{tabularx}
\usepackage{booktabs}
\usepackage{multirow}

\begin{document}

\title{
Directional Cluster Migration Driven by Escape-Rate Asymmetry in Multi-Compartment Granular Systems
}

\author{Kai Kono}
\affiliation{Department of Physics, Graduate School of Science, Kyushu University, Fukuoka, Japan}

\author{Hiroyuki Ebata}
\affiliation{Department of Physics, Graduate School of Science, Kyushu University, Fukuoka, Japan}
\affiliation{Department of Earth and Space Science, The University of Osaka, Toyonaka, Japan}

\author{Shio Inagaki}
\email{shio_inagaki@sis.u-hyogo.ac.jp}
\affiliation{Department of Physics, Graduate School of Science, Kyushu University, Fukuoka, Japan}
\affiliation{Graduate School of Information Science, University of Hyogo, Kobe, Japan}

\date{\today}

\begin{abstract}
Granular materials are inherently out-of-equilibrium systems due to energy dissipation through inelastic collisions and friction. When driven by mechanical agitation such as vibration, they exhibit rich collective behaviors including segregation, clustering, and spontaneous oscillations. Here, we report directional stepwise migration of particle clusters from one compartment to the next in a vertically vibrated granular system composed of small and large particles. To clarify the underlying mechanism, we directly measured how the flux of both particle species depends on the instantaneous particle populations. The measurements reveal an asymmetric interaction between particle species: the flux of small particles is enhanced by the presence of large particles, whereas that of large particles is suppressed by small particles. A minimal flux model incorporating these measured fluxes reproduces the observed directional dynamics and provides an experimentally grounded framework for collective transport in vibrated granular systems.
\end{abstract}

\maketitle
\section{Introduction}
Granular materials are inherently out-of-equilibrium systems in which energy is dissipated through inelastic collisions and friction between macroscopic particles. When mechanical energy is continuously supplied through vibration or rotation, these dissipative interactions can give rise to nontrivial collective behavior, including spontaneous symmetry breaking and the formation of inhomogeneous particle distributions. Examples include size segregation, where particles of different sizes separate spontaneously \cite{Aranson:_PatternFormation,Ottino:_SegregationReview}, and clustering, where particles aggregate into dense regions despite uniform energy input \cite{Goldhirsch:_GranularGas,Kudrolli1997,vdWeele:_Review}.

In monodisperse systems, the phenomenon known as "Sand as Maxwell's Demon" has been reported \cite{Eggers:_SandDemon}. In this setup, a container is divided into two compartments connected by a narrow slit near the bottom and partially filled with identical particles that are initially equally distributed. Under vertical vibration, once the driving strength exceeds a certain threshold, the symmetric state becomes unstable and particles spontaneously cluster into one compartment. This transition has been theoretically described as a pitchfork bifurcation, based on the particle escape rate from a compartment derived using kinetic theory \cite{Lohse:_MonoBifurcation,vanderWeele2001hysteretic}. Beyond two-compartment systems, more complex spatiotemporal dynamics emerge in circular and multi-compartment geometries \cite{Lohse2004:_Coarsening,Lohse2002:_Collapse,Lohse2004:_Ratchet}. Experiments have reported coarsening phenomena, where clusters merge over time \cite{Lohse2004:_Coarsening}, as well as abrupt collapse events in which clusters suddenly dissolve into a gas-like state \cite{Lohse2002:_Collapse}.

Extending the two-compartment system to bidisperse mixtures leads to qualitatively different dynamics. In two-dimensional numerical simulations, Lambiotte {\it et al.} \cite{Lambiotte2005:_GranularClockSIM} discovered a striking oscillatory state in which particle clusters spontaneously shuttle between two compartments, which they termed the "granular clock". This behavior was later confirmed experimentally by Viridi {\it et al.} \cite{Viridi:_GranularClockEX}.
Oscillatory dynamics has been reported in systems where particles differ either in size at equal density \cite{Viridi:_GranularClockEX} or in density at equal size \cite{MHou2008:_Temperature}. In three-compartment systems, stochastic switching of clusters between compartments has been observed \cite{KCChen2008:_3Cells}. In five-compartment systems, recurrent stepwise cluster motion across neighboring compartments has been reported \cite{Liu2008:_ThreeCompartments}. However, these experimental reports remain largely qualitative, and directional stepwise migration of clusters from one compartment to the next has not been quantitatively characterized in multi-compartment systems.

To describe the dynamics of such compartmentalized granular systems, flux-based models have frequently been employed. In these models, the time evolution of the particle population in each compartment is governed by escape rates that characterize transport between neighboring compartments. In monodisperse systems, such escape-rate functions were originally introduced on the basis of kinetic-theory arguments for vertically vibrated granular gases \cite{Eggers:_SandDemon}. Subsequent studies extended this framework to bidisperse mixtures by introducing phenomenological flux functions that incorporate segregation effects and reproduce oscillatory cluster dynamics \cite{Lambiotte2005:_GranularClockSIM,Viridi:_GranularClockEX}. Despite their success in capturing qualitative features of the dynamics, the escape-rate functions used in these models have generally been assumed or derived under simplifying theoretical approximations rather than measured directly in experiments.

\begin{figure}[bhtp]
  \begin{center}
  \includegraphics[width=\columnwidth]{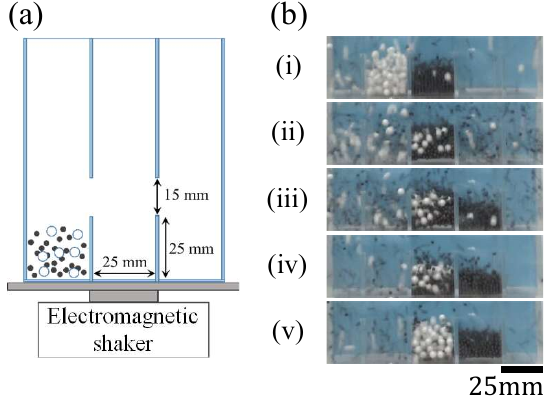}
  \end{center}
  \caption{
  (a) Schematic illustration of the experimental setup (front view). The cell depth, perpendicular to the plane of the figure, was fixed at $30 \,\rm{mm}$. (b) Representative sequential snapshots taken at $10 \,\rm{s}$ intervals, showing the cluster advancing by one compartment in a five-compartment cell ($K=5$) with $N^S=1350$ and $N^L=124$.
  }
  \label{fig:1}%
\end{figure}
In this study, we demonstrate persistent directional migration of a particle cluster in vertically vibrated granular systems with up to seven compartments. To explain this behavior, we focus on the flux dynamics in bidisperse mixtures and develop a minimal flux-based model grounded entirely in experimentally measured escape-rate functions. By systematically varying the number of small and large particles in a compartment, we directly measure the corresponding particle flux using high-speed video recordings. Here the particle flux is defined as the number of particles that escape from a compartment per unit time. These measurements reveal a clear asymmetric interaction between particle species: the escape rate of small particles is enhanced by the presence of large particles, whereas the escape rate of large particles is suppressed by small particles. Incorporating these experimentally determined escape rates into the flux model allows us to reproduce the observed directional cluster motion without introducing additional phenomenological coupling terms between particle species. 
This asymmetric coupling is reminiscent of activating-inhibitory interactions in reaction-diffusion systems \cite{Turing:_ReacDiff,Meinhardt:_ReacDiffReview,Murray2003} and related population-dynamics models such as the Lotka-Volterra equations \cite{Goel1971}.

\section{Collective dynamics in multiple compartments}
The experimental setup is shown in Fig. \ref{fig:1}(a). 
We used black silicon nitride spheres (diameter $d_S= 2 \,\rm{mm}$, mass density of $3.1 \,\rm{g/cm^3}$) and white zirconia spheres (diameter $d_L=4 \,\rm{mm}$, mass density of $6.0 \,\rm{g/cm^3}$). We independently measured the restitution coefficients to be $e_S=0.91 \pm 0.01$ for the small particles and  $e_L=0.86 \pm 0.02$ for the large particles. The cell was constructed from acrylic plates, with a height of $150 \,\rm{mm}$ and a front-to-back depth of $30 \,\rm{mm}$. The compartments were separated by acrylic dividers spaced at $25 \,\rm{mm}$ intervals. Each divider had a 15-mm-wide opening, with its lower edge located $25\,\rm{mm}$ above the bottom, allowing particles to move between compartments. The thickness of each divider was $2 \,\rm{mm}$. The number of compartments, $K$, was varied from 2 to 7, so the total width of the container depended on $K$. 
We also varied the numbers of small and large particles, $N^S$ and $N^L$, to examine how the collective dynamics depend on the particle composition.
The cell was sealed at the top and mounted on a sinusoidal electromagnetic shaker (EMIC 512-A). 
Throughout the experiments, the vibration frequency $f$ and amplitude $a$ were fixed at $30 \,\rm{Hz}$ and $2 \,\rm{mm}$, respectively. With these parameters, the dimensionless acceleration was  $\Gamma = (2\pi f)^2 a / g \approx 7.1$, where $g$ denotes the gravitational acceleration. As an initial condition, the two types of particles were mixed and placed into the leftmost compartment before the vibration started. To ensure reproducibility, particles were always placed in the same compartment at the beginning of each experiment.

A video camera was used to record the system at 30 frames per second. To visualize the cluster dynamics, spatiotemporal diagrams were constructed by vertically stacking one-pixel-high horizontal lines, where each line corresponds to a single time frame. For each frame, the RGB values were averaged over the vertical direction from the bottom of the cell up to the slit height ($25 \,\rm{mm}$), yielding a one-pixel-high horizontal profile along the compartment array. 
In the spatiotemporal plots, the time origin was defined such that $t=0$ corresponds to the onset of vibration.

\begin{figure}[thbp]
  \begin{center}
  \includegraphics[width=\columnwidth]{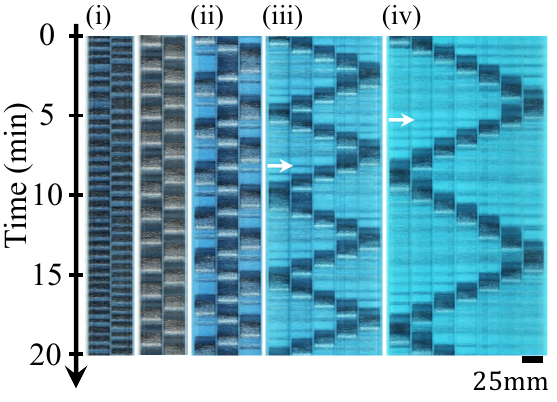}
  \end{center}
  \caption{
  Spatiotemporal diagrams for systems with $K=2, 3, 5,$ and $7$. Time  $t=0$ corresponds to the onset of vibration. (i) $K=2$. Two parameter sets are shown: $N^S=347$ and $N^L=25$ (left), and $N^S=1000$ and $N^L=100$ (right). (ii-iv) $K=3, 5,$ and $7$. In all cases, $N^S=1350$ and $N^L=124$. In the spatiotemporal diagrams, black and white represent small and large particles, respectively. Arrows indicate faint horizontal streaks corresponding to gas-like small particles. $N^S$ and $N^L$ denote the numbers of small and large particles, respectively.}
  \label{fig:2}
\end{figure}

\subsection{Directional cluster migration across multiple compartments}
Within a certain range of particle numbers, the bidisperse granular system exhibits directional cluster migration, in which a localized cluster advances stepwise from one compartment to the next under vertical vibration.
To aid visualization, representative snapshots of the cluster motion in a five-compartment system are shown in Fig. \ref{fig:1}(b). We first characterize this behavior systematically using the spatiotemporal diagrams in Fig. \ref{fig:2} for $K=2, 3, 5,$ and $7$.
Corresponding movies are provided in Supplementary Materials A (Movies A1-A5).

Beginning with the well-studied two-compartment system ($K=2$), we found two oscillatory states with distinct migration timescales by varying only the particle numbers (Fig. \ref{fig:2}(i)). For $N^S=347$ and $N^L=25$, the average time required for the cluster to move from one compartment to the other is approximately  $20 \,\rm{s}$, whereas for $N^S=1000$ and $N^L=100$, it increases to approximately $48 \,\rm{s}$.

In the three-compartment system (Fig. \ref{fig:2}(ii)), we observed periodic oscillation of the cluster. It moved successively from the left compartment to the center, then to the right, and back again.
This indicates that the cluster migration proceeds stepwise through successive transitions between neighboring compartments.
By contrast, a previous study with the same number of compartments reported stochastic cluster motion \cite{KCChen2008:_3Cells}. One possible origin of this discrepancy is the front-to-back depth of the cell. Although the same particle diameters ($2 \,\rm{mm}$ and $4 \,\rm{mm}$) were used in that study, the cell depth was only about $7 \,\rm{mm}$, less than one-fourth of that in our setup. In our experiments, stepwise periodic cluster motion was already observed in a three-compartment cell with a depth of $10 \,\rm{mm}$, but remained intermittently unstable. Increasing the cell depth to $30 \,\rm{mm}$ made the migration markedly more regular and was essential for achieving stable directional migration in systems with larger $K$.

For $K=5$ and $7$, the cluster exhibits sustained periodic back-and-forth migration across the compartments, reversing direction at the system boundaries (Fig. \ref{fig:2}(iii) and (iv)). Notably, to our knowledge, such robust directional cluster motion in a seven-compartment system has not been reported previously. 

Figure \ref{fig:1}(b) presents a representative sequence of snapshots illustrating one cycle of directional migration. The sequence begins with a separated configuration that forms spontaneously during migration. In panel (i), the large-particle cluster is located immediately to the left of the small-particle-rich compartment. Because large particles tend to be trapped in the small-particle-rich compartment through dissipative collisions with small particles, the large-particle cluster gradually shifts to the right. As large particles accumulate in that compartment, the escape of small particles into the next compartment is enhanced. This process is visible in panels (iii) and (iv), where the small-particle-rich region is pushed forward. By panel (v), the two species are again spatially separated into neighboring compartments but shifted by one compartment relative to panel (i). Repetition of this sequence produces the persistent directional migration of the cluster. A similar chasing behavior has also been reported in a three-compartment system\cite{KCChen2008:_3Cells}.

Although Fig. \ref{fig:1}(b) appears to show the large-particle cluster chasing the small-particle cluster, the observed directionality emerges only at the collective level. Individual particles, however, move approximately symmetrically to the left and right. A closer look at panels (iii)-(v) reveals that large white particles are also present in a gas-like, spatially dispersed state in the compartments to the left of the large-particle-rich compartment. This interpretation is also consistent with the spatiotemporal diagrams in Fig. \ref{fig:2}, where faint horizontal dark streaks (indicated by arrows) extend across multiple compartments. These streaks correspond to small particles in a gas-like state and reflect their approximately symmetric emission toward neighboring compartments. Large particles are also emitted approximately symmetrically, although their contribution is less clearly visible in the spatiotemporal diagrams. 

\begin{figure}[t]
  \begin{center}
  \includegraphics[width=\columnwidth]{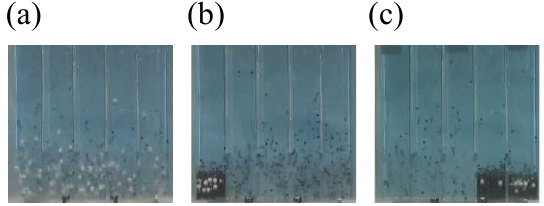}
  \end{center}
  \caption{
 Representative snapshots ($K=5$) illustrating three selected states used for phase-diagram classification. (a) Gas-like state. (b) Single-cluster state. (c) Two-cluster state. The directional migration state is shown separately in Fig. \ref{fig:1}(b).
  }
  \label{fig:3}%
%
  \begin{center}
  \includegraphics[width=0.9\columnwidth]{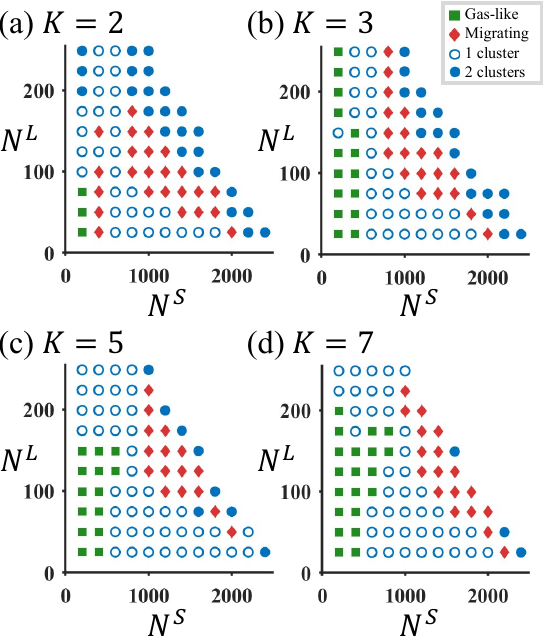}
  \end{center}
  \caption{
Phase diagrams classifying the dynamical states of the clusters in the $N^S-N^L$ plane. The number of compartments is (a) $K=2$, (b) $K=3$, (c) $K=5$, and (d) $K=7$. Green squares: gas-like state; red diamonds:  migrating state; blue open circles: stationary single-cluster state; blue filled circles: stationary two-cluster state.
  }
  \label{fig:4}%
\end{figure}

\subsection{Phase diagrams}
By systematically varying the numbers of small and large particles, $N^S$ and $N^L$, we identified three additional dynamical states besides the directional migration state shown in Fig. \ref{fig:2}.  Representative snapshots of the three stationary states are shown in Fig. \ref{fig:3}. Phase diagrams summarizing all four states were then constructed in the $N^S-N^L$ plane for $K=2, 3, 5,$ and $7$ (Fig. \ref{fig:4}). In all cases, at very low total particle numbers, particles remained in a dilute, gas-like state without forming a persistent cluster (Fig. \ref{fig:3}(a)). As the particle numbers increased, a stationary single-cluster state emerged  (Fig. \ref{fig:3}(b)). In this state, increased energy dissipation through inelastic collisions suppressed gas-like dispersion, and the particles remained trapped as a localized cluster in the initially occupied leftmost compartment. At higher particle numbers, a stationary two-cluster state was observed  (Fig. \ref{fig:3}(c)). Typically, the migrating cluster reached the right boundary but could not reverse its motion, resulting in two localized clusters trapped in the two rightmost compartments. Corresponding movies of the gas-like, stationary single-cluster, and stationary two-cluster states in Fig. \ref{fig:3} are provided in Supplementary Materials A (A6-A8). 

An intermediate dynamical region was observed between the single- and two-cluster regimes, where the cluster underwent spontaneous directional migration (Fig. \ref{fig:4}). Near the phase boundaries, however, the motion sometimes became irregular or stochastic.
For $K=2$, this migrating state appeared as oscillatory motion between the two compartments.
In addition, a narrow oscillatory region was also found between the gaseous and single-cluster states. This lower-total-particle region corresponds to Fig. \ref{fig:2}(i)(left), whereas the higher-particle-number region corresponds to Fig. \ref{fig:2}(i)(right). In the lower-particle-number region, oscillations were observed over a relatively wide range of $N^L$, whereas the allowable range of $N^S$ was much narrower, indicating that $N^S$ acted as a key control parameter for the dynamics in this regime. 

As the number of compartments increased, the migrating region shifted toward the upper right in the $N^S-N^L$ plane, indicating that a larger total number of particles was required to sustain cluster migration for larger $K$. Interestingly, the migrating region exhibited a downward-sloping shape, suggesting that migratory behavior was more likely to occur when the total number of particles fell within a certain range, rather than being determined primarily by the ratio of small to large particles. The region where both $N^S$ and $N^L$  were large remained unexplored because the total number of particles exceeded the capacity of the initially occupied leftmost compartment. 

In the case of $K=9$, the migrating region is expected to become even narrower. Using the same particle numbers as in Fig. \ref{fig:2}(ii-iv), the cluster completed only a single forward-and-return cycle at a lower vibration strength ($\Gamma=5.2$; figure not shown). This was the longest directional motion observed so far in the nine-compartment system. These results suggest that directional cluster motion may still be achievable in systems with more than seven compartments, provided that the particle numbers and vibration strength are carefully tuned.

\vspace{5mm}
\begin{figure}
  \begin{center}
  \includegraphics[width=0.7\columnwidth]{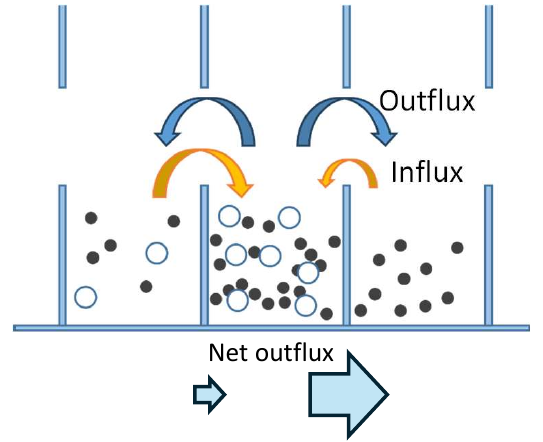}
  \end{center}
  \caption{
Schematic illustration of the local particle balance that determines the direction of cluster migration.
  }
  \label{fig:5}%
\end{figure}
\begin{figure}
  \begin{center}
  \includegraphics[width=0.9\columnwidth]{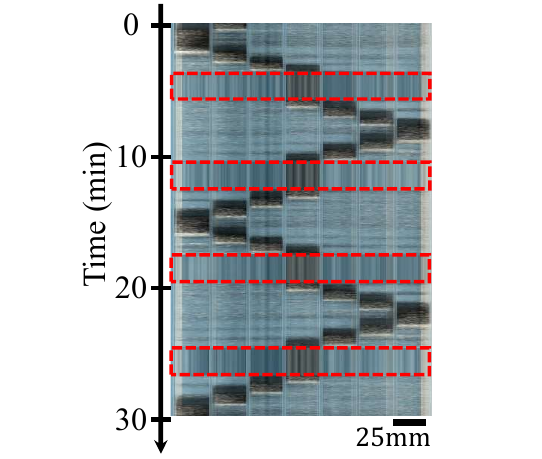}
  \end{center}
  \caption{
Spatiotemporal diagram for a seven-compartment cell, with the same particle numbers as in Fig. \ref{fig:2}(d). The vertical vibration was interrupted four times for two minutes each. Red dashed boxes indicate intervals without vibration.
  }
  \label{fig:6}%
\end{figure}

\subsection{Local particle flux balance and migration direction}
A key question is what determines the direction of cluster motion. Figure \ref{fig:5} schematically illustrates the local particle balance around a migrating cluster. For a localized cluster to migrate while maintaining its structure, the local balance between particle influx and outflux must differ between the forward and backward sides of the cluster. On the forward side, the outflux from the cluster-containing compartment must exceed the influx from the neighboring compartment so that the cluster can advance. On the backward side, by contrast, the influx and outflux must be comparable, or the influx may even exceed the outflux, so that the localized cluster is not depleted too rapidly. Because the geometry is left-right symmetric, particle escape from the cluster-containing compartment is expected to be nearly symmetric toward the two neighboring compartments. The migration direction is therefore determined mainly by the imbalance in the particle influxes from those neighboring compartments. The cluster thus advances toward the side with the smaller influx, where the net particle outflux is larger.

\begin{figure*}
  \begin{center}
  \includegraphics[width=\textwidth]{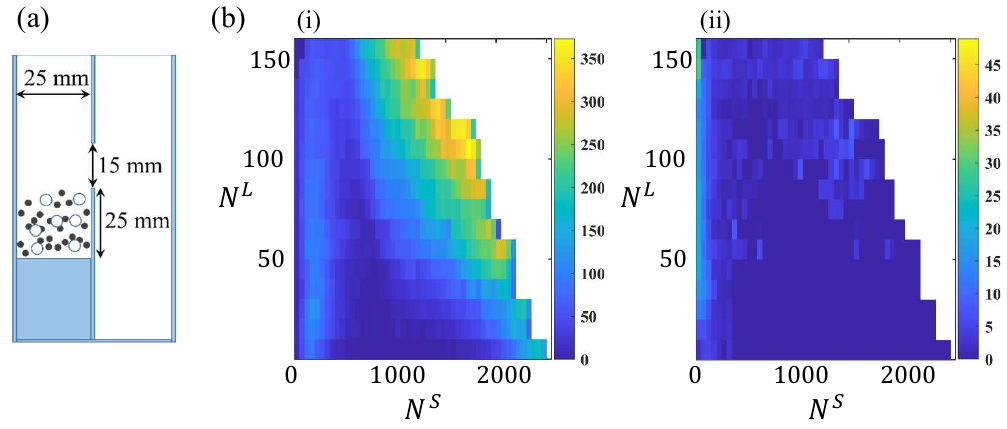}
  \end{center}
  \caption{
(a) Asymmetric cell to measure particle escape rate. (b) Heat maps of the escape rates: (i) small particles, $F^S (N^S,N^L)$; (ii) large particles, $F^L (N^S,N^L)$. The maps were constructed by averaging the measured data over intervals of 50 particles in $N^S$ and 10 particles in $N^L$. 
}
  \label{fig:7}
\end{figure*}

To test whether the migration direction is determined by the instantaneous particle distribution or by the prior trajectory of the system, we performed an intermittent driving experiment. Figure \ref{fig:6} shows the spatiotemporal diagram of a seven-compartment system when the vibration was temporarily halted and then restarted. Although the cluster motion ceased while the vibration was off, the same migration direction was recovered once the vibration was resumed. This indicates that the escape rate of each species depends on the instantaneous populations of small and large particles within each compartment. Taken together, these results suggest that both the migration direction and the ability to maintain a localized cluster are governed by the instantaneous particle populations rather than by the prior history.

\section{Population-dependent escape rates}
Motivated by these observations, we directly measured how the escape rates of small and large particles depend on the instantaneous particle populations within a compartment.
We then incorporated these experimentally determined escape rates into a minimal flux model, in which the particle numbers in each compartment evolve through net fluxes between neighboring compartments.
This approach avoids introducing phenomenological escape-rate functions based on kinetic-theory assumptions.

\subsection{Direct measurement of escape rates}
For this purpose, 
we designed an asymmetric cell in which the height of the left compartment was raised to prevent particles from reentering from right to left (Fig. \ref{fig:7}(a)).
The divider opening was $15 \, \rm{mm}$ wide and began $25 \, \rm{mm}$ above the bottom of the left compartment. The dimensions of the left compartment and the thickness of the divider were identical to those of the granular clock cell described above.
A high-speed camera (HAS-U2, DITECT) operating at $800 \, \rm{fps}$ was used to record particle motion near the slit. Various combinations of small and large particles were placed in the left compartment, and vibration was applied under the same conditions as in the main experiments. We counted the numbers of small and large particles crossing the slit from left to right. The escape rates were obtained by differentiating the cumulative number of escaping particles with respect to time while simultaneously tracking the instantaneous numbers of small and large particles remaining in the left compartment.
As particles exited the left compartment, $N^S$ and $N^L$ changed over time. To cover a wide range of particle numbers, experiments were performed with multiple initial combinations. 
The measurements from different initial conditions collapse onto the same functional dependence (see Supplementary Materials B). Together with the behavior observed in Fig. \ref{fig:6}, these results indicate that the escape rates are governed primarily by the instantaneous particle populations rather than by the initial preparation.
Each experimental condition was repeated 10 times, and the escape rates were obtained by averaging over these independent realizations.

Figure \ref{fig:7}(b) presents heat maps of the escape rates of small and large particles, providing a global view of how the escape dynamics depend on the instantaneous particle populations. For small particles, the escape rate exhibits a prominent ridge at high $N^S$, together with a weaker ridge at low $N^S$ (Fig. \ref{fig:7}(b-i)). For large particles, the escape rate remains nearly zero over most of the parameter range, except at low $N^S$ (Fig. \ref{fig:7}(b-ii)). This indicates a strong asymmetry in the escape dynamics, where small particles escape preferentially when both species coexist in a compartment, while the escape of large particles is effectively suppressed until the number of small particles has been sufficiently reduced. This trend is consistent with the behavior observed in Fig. 1b. The heat maps reveal smooth trends, indicating that the escape rates vary systematically with the particle numbers. 

To examine the functional dependence more directly, we constructed one-dimensional profiles of the escape rates as functions of particle number. To reduce statistical fluctuations, the data were averaged over finite ranges of the complementary particle number. For the small-particle escape rates, relatively narrow ranges were sufficient ($N^L \pm 1$; Fig. \ref{fig:8}(a)). For the large-particle escape rates, wider averaging ranges were used because escape events were much less frequent, especially at large $N^S$, where small particles preferentially escaped from the compartment. Specifically, we used $N^S \pm 10$ for $N^S \approx 10$, and $N^S \pm 20$ for $N^S \approx 40$ and $120$ (Fig. \ref{fig:8}(b)). The curves for $N^S \approx 40$ and $N^S \approx 120$ were obtained from the bidisperse dataset used for the heat map in Fig. \ref{fig:7}(b-i), and therefore covered $0 \le N^L \lesssim 40$. To probe the high-$N^L$ regime, we also included the $N^S = 0$ profile, obtained by combining the bidisperse data at $N^S = 0$ with independent monodisperse measurements.

Figure \ref{fig:8}(a) shows the escape rate of small particles as a function of $N^S$ for different values of $N^L$. Across all $N^L$, the curves exhibit a characteristic peak at low $N^S$, followed by a decrease and then a rise at higher $N^S$.
In the low-$N^S$ regime, the escape rate increases with particle number because a larger population leads to more frequent escape attempts. 
As $N^S$ increases further, more frequent interparticle collisions enhance energy dissipation and temporarily reduce the escape rate. At sufficiently large $N^S$, the height of the granular layer increases, effectively reducing the available slit height and leading to overflow and a sharp rise in the escape rate. The escape rate of small particles increases with $N^L$ while preserving the overall shape of the curves. The experimentally accessible range of $N^S$ decreases with increasing $N^L$ because large particles occupy part of the compartment volume.

Figure \ref{fig:8}(b) shows the escape rate of large particles as a function of $N^L$ for different values of $N^S$. For $N^S \approx 10$, the large-particle escape rate exhibits a peak-and-rise structure similar to that observed for small particles. As the number of small particles increases, however, the low-$N^L$ peak decreases in magnitude. This trend is consistent with the idea that collisions with an increasing number of small particles enhance energy dissipation and thereby suppress the escape of large particles.

These results reveal a clear asymmetric interaction between the two particle species. Large particles promote the escape of small particles, whereas small particles suppress the escape of large particles. This asymmetric coupling naturally explains the preferential release of small particles and provides a key ingredient for the directional cluster migration observed in Fig. \ref{fig:1}(b).

\subsection{Functional forms of the escape rates}
\begin{figure}[htbp]
  \begin{center}
  \includegraphics[width=\columnwidth]{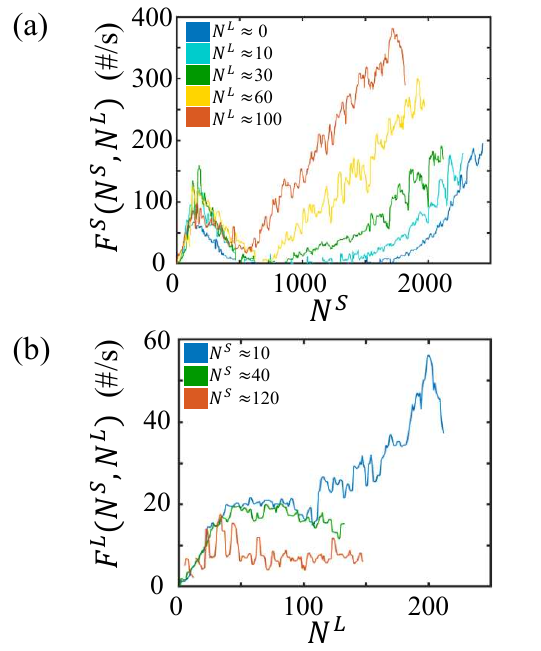}
  \end{center}
  \caption{
Escape rates: (a) small particles, $F^S (N^S,N^L)$ for $N^L \approx 0,10,30,60,$ and $100$; (b) large particles, $F^L (N^S,N^L)$ for $N^S \approx 10,40,$ and $120$.
  }
  \label{fig:8}
\end{figure}

\begin{figure}[htbp]
  \begin{center}
  \includegraphics[width=\columnwidth]{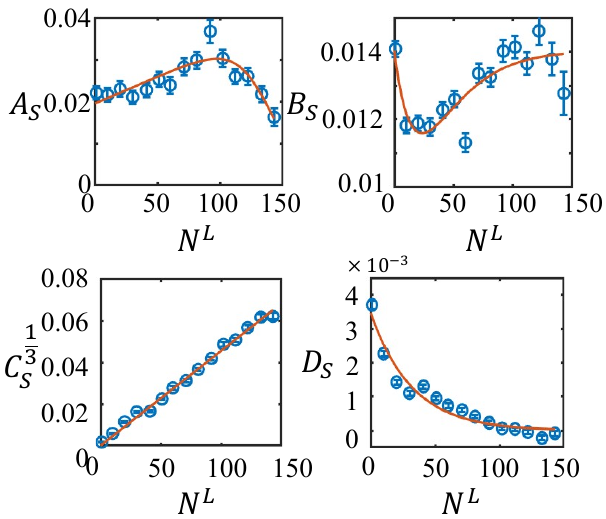}
  \end{center}
  \caption{
 Fitting parameters $A_S, B_S, C_S$, and $D_S$ for the small-particle escape rate $F^S (N^S,N^L )$ as functions of  $N^L$. Symbols represent the fitted values obtained for each $N^L$, and solid lines show the fitting curves described by Eqs. (2)-(5). The error bars represent $95 \%$ confidence intervals of the fitting. The fitted coefficients are $\alpha_{S1}=(6.5 \pm 4.2)\times 10^(-5)$,
 $\alpha_{S2}=(9.8 \pm 1.9)\times 10^(-3)$,
 $\alpha_{S3}=(3.4 \pm 2.6) \times 10^(-2)$,
 $\alpha_{S4}=(1.38 \pm 0.09)\times 10^2$, 
 $\beta_{S1}=(2.8 \pm 1.6) \times 10^(-4)$,
 $\beta_{S2}=(4.3 \pm 1.7) \times 10^(-2)$,
 $\beta_{S3}=(1.40 \pm 0.07) \times 10^(-2)$, 
 $\gamma_{S1}=(9.4 \pm 0.7) \pm 10^(-11)$, 
 $\delta_{S1}=(3.5 \pm 0.5) \times 10^(-3)$,
 $\delta_{S2}=(3.1 \pm 0.7) \times 10^(-2)$.
  }
  \label{fig:9}
\end{figure}

To quantitatively incorporate the observed asymmetric interaction into the flux model, we determined empirical functional forms for the escape rates of small and large particles. For small particles, the one-dimensional profiles of the escape rate as a function of $N^S$ at fixed values of $N^L$ (representative examples are shown in Fig. \ref{fig:8}(a)) were fitted using
\begin{equation}
  \begin{aligned}
    F^S(N^S, N^L)
    &= {N^S}^{2}\,
       \left[
         A_S\,\exp \! \bigl(-B_SN^S\bigr)
       \right.
       \\[4pt]
    &\quad
       \left.
         +\,C_S\,\exp \! \bigl(D_SN^S\bigr)
       \right]
       ,\label{eq:flux_S}
\end{aligned}
\end{equation}
where $A_S$, $B_S$, $C_S$, and $D_S$ are fitting parameters determined separately for each fixed $N_L$ (see Supplementary Materials B for representative fits).
The fitting was performed for many closely spaced values of $N^L$ across the dataset. The first term corresponds to the dilute regime and captures the peak at low $N^S$.
This form is motivated by the kinetic-theory expression for dilute gases, $F(N) \propto N^2  \exp(-\alpha N^2 )$ \cite{Eggers:_SandDemon}.
In our system, a modified form $F(N) \propto N^2  \exp(-\alpha N)$ provided better agreement with the experimental data while preserving the essential peaked structure with rapid decay (see Supplementary Materials B for a comparison with $N^2  \exp(-\alpha N^2))$.
For each fixed $N^L$, the parameters were obtained by nonlinear least-squares fitting in MATLAB.
The fitted parameters vary smoothly and systematically with  $N^L$, as shown in Fig. \ref{fig:9}.
This dependence is well described by
\begin{align}
  A_S(N^L) &= (\alpha_{S1} N^L + \alpha_{S2})(1-\tanh(\alpha_{S3}(N^L-\alpha_{S4}))), \label{eq:sa}
  \\
  B_S(N^L) &= (\beta_{S1} N^L + \beta_{S2})\exp(-\beta_{S3} N^L)-\beta_{S4},\label{eq:sb}
  \\
  C_S(N^L) &= \gamma_{S1} {N^L}^3\label{eq:sc},\\
  D_S(N^L) &= \delta_{S1} \exp(-\delta_{S2} N^L),\label{eq:sd}
\end{align}
where the Greek coefficients are fitting constants.
These empirical forms were introduced to reproduce the overall shape of the data and are not unique.
While $A_S(N^L)$ exhibits a slight decrease at large $N^L$, it increases over the moderate-$N^L$ range, whereas $C_S(N^L)$ increases monotonically with $N^L$.
Because the contribution of the first term becomes less significant at large $N^L$, these fitted functions consistently indicate that the escape rate of small particles is enhanced by the presence of large particles.
Substituting Eqs.~(\ref{eq:sa})--(\ref{eq:sd}) into Eq.~(\ref{eq:flux_S}) yields the final functional form for $F^S(N^S, N^L)$.

\begin{figure}[htbp]
  \begin{center}
  \includegraphics[width=0.9\columnwidth]{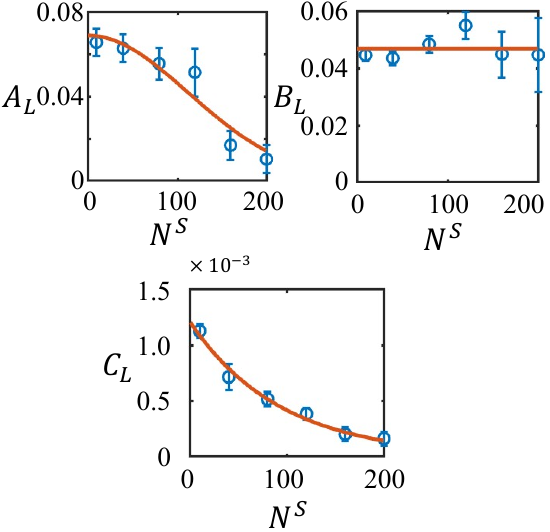}
  \end{center}
  \caption{
Fitting parameters $A_L, B_L$, and $C_L$ for the large-particle escape rate $F^L (N^S,N^L)$ as functions of  $N^S$. Symbols represent the fitted values obtained for each $N^S$, and solid lines show the fitting curves described by Eqs. (7)-(9). The error bars represent $95 \%$ confidence intervals of the fitting. The fitted coefficients are 
$\alpha_{L1}=(6.9 \pm 1.5) \times 10^(-2)$,
$\alpha_{L2}=(4.0 \pm 2.5) \times 10^(-3)$,
$\beta_{L1}=(4.7 \pm 0.5) \times 10^(-2)$,
$\gamma_{L1}=(1.2 \pm 0.2) \times 10^(-3)$,
$\gamma_{L2}=(1.1 \pm 0.2) \times 10^(-2)$.
}
\label{fig:10}%
\end{figure}

The escape rate of large particles was analyzed following the same procedure. We adopted a fitting function analogous to Eq.~(\ref{eq:flux_S}), but simplified due to the more moderate variation in the escape rate:
\begin{equation}
  F^L(N^S, N^L) = {N^L}^{2}\,
     \left[
      A_L\,\exp \! \bigl(-B_LN^L\bigr)+\,C_L
      \right].\label{eq:flux_L}
 \end{equation}
Here, $A_L$, $B_L$, and $C_L$  are fitting parameters determined separately for each fixed $N^S$. The exponential term captures the initial growth followed by decay typical of low to moderate $N^L$, while the constant term accounts for a baseline contribution at high $N^L$. For each fixed $N^S$, the parameters were obtained by nonlinear least-squares fitting in MATLAB. The fitting was performed for several values of $N^S$ across the dataset. The fitted parameters exhibit systematic and smooth dependence on $N^S$, as shown in Fig. \ref{fig:10}. This dependence is well described by
\begin{align}
  A_L(N^S) &= \alpha_{L1} \exp(-\alpha_{L2} {N^S}^2) \label{eq:la},\\
  B_L(N^S) &= \beta_{L1}, \label{eq:lb}\\
  C_L(N^S) &= \gamma_{L1} \exp(-\gamma_{L2} N^S),\label{eq:lc}
\end{align}
where the Greek coefficients are fitting constants. Unlike the case for small particles, both $A_L(N^S)$ and $C_L(N^S)$ decrease with increasing $N^S$, indicating that the presence of small particles suppresses the escape of large particles at fixed $N^L$. Substituting Eqs.~(\ref{eq:la})--(\ref{eq:lc}) into Eq.~(\ref{eq:flux_L}) yields the final functional form for $F^L (N^S,N^L)$.
Together with the corresponding expression for $F^S(N^S,N^L)$, this completes the experimentally determined model for the population-dependent escape rates.


\section{Flux model}
\begin{figure}[t]
  \begin{center}
  \includegraphics[width=\columnwidth]{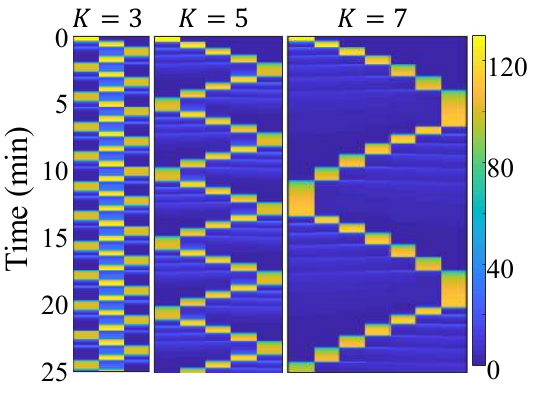}
  \end{center}
  \caption{
Spatiotemporal color maps of $N^L$ in each compartment obtained from numerical solutions of the flux model. The fitted coefficients obtained in Figs. 8 and 9 were used in Eqs. (2)-(5) and (7)-(9). The initial conditions are $N_1^S=1400$ and $N_1^L=130$, with $N_i^S=N_i^L=0$ ($i>1$).
}
\label{fig:11}
\end{figure}

Using these experimentally determined escape rates, we constructed a flux model to describe the collective dynamics of the system. Assuming that particles migrate only between adjacent compartments and that escape from interior compartments occurs symmetrically to the left and right, the time evolution of the numbers of small and large particles
in the $i$-th compartment, $N^S_i$ and $N^L_i$, is given by
\[
\left\{
\begin{aligned}
  \frac{dN^S_i}{dt} &=-2\,F^S\!\left(N^S_i, N^L_i\right) + \sum_{j=\pm 1} F^S\!\left(N^S_{i+j}, N^L_{i+j}\right), \\[4pt]
  \frac{dN^L_i}{dt} &= -2\,F^L\!\left(N^S_i, N^L_i\right) + \sum_{j=\pm 1} F^L\!\left(N^S_{i+j}, N^L_{i+j}\right),
\end{aligned}
\right.
\]
where $i=1,...,K$, and $K$ is the total number of compartments \cite{Lohse2007:_Review,Lambiotte2005:_GranularClockSIM,Viridi:_GranularClockEX}.
To impose no-flux boundary conditions, we set $N^S_0 = N^S_1$, $N^L_0 = N^L_1$, $N^S_{K+1} = N^S_K$, and $N^L_{K+1} = N^L_K$,  so that the edge compartments exchange particles only with their single neighboring compartment. This formulation yields a system of $2K$ coupled ordinary differential equations, which can be solved deterministically for given initial conditions. The equations were numerically integrated using the Euler method, employing the experimentally determined functional forms of $F^S(N^S,N^L )$ and $F^L(N^S,N^L )$.
As in the experiments, the initial condition was chosen such that $N_1^S=1400$ and $N_1^L=130$, while $N_i^S=0$ and $N_i^L=0$ for $2 \le i \le K$. Figure \ref{fig:11} presents representative numerical spatiotemporal patterns for $K=3, 5$, and $7$, shown as color maps of  $N^L$. For systems with up to seven compartments, the flux-based model successfully reproduces the sustained directional cluster motion observed in experiments. The model considers only particle exchange between adjacent compartments and uses experimentally measured escape rates, without introducing additional phenomenological couplings. This demonstrates that the essential mechanism underlying directional migration is captured by escape rates that depend explicitly on both $N^S$ and $N^L$.

\begin{figure}[htbp]
  \begin{center}
  \includegraphics[width=\columnwidth]{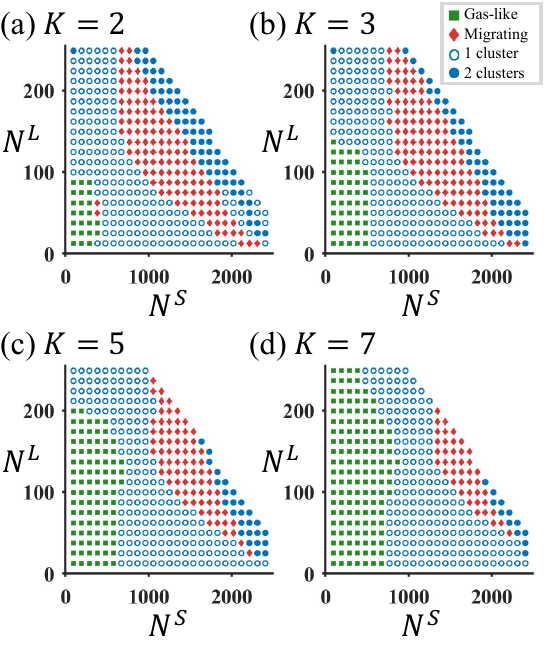}
  \end{center}
  \caption{
Phase diagrams classifying the dynamical states of the clusters in the $N^S-N^L$ plane, obtained from numerical solutions of the flux model. The number of compartments is (a) $K=2$, (b) $K=3$, (c) $K=5$, and (d) $K=7$. Green squares: gas-like state; red diamonds:  migrating state; blue open circles: stationary single-cluster state; blue filled circles: stationary two-cluster state.
}
\label{fig:12}
\end{figure}

Following the experimental procedure, we varied the numbers of small and large particles and numerically solved the flux model for each combination. The resulting dynamical state was classified for each parameter set, and phase diagrams were constructed in the same manner as in the experiments. Figure \ref{fig:12} presents the corresponding phase diagrams, directly comparable to the experimental phase diagrams shown in Fig. \ref{fig:4} for $K=2, 3, 5$, and $7$. The migrating region lies between the single- and two-cluster regimes. As the number of compartments $K$ increases, the migrating region becomes narrower, and a larger total number of particles is required to sustain directional cluster motion. For $K=2$, an additional narrow migrating region appears at low total particle numbers, consistent with the experimental phase diagram (Fig. \ref{fig:4}(a)). For consistency with the experimental protocol, the region where both $N^S$ and $N^L$ are large is also excluded from the numerical phase diagrams, because the total number of particles exceeds the capacity of the leftmost compartment under the prescribed initial condition.

Overall, the model reproduces the main structural features of the experimental phase diagrams, including the two distinct migrating regions for $K=2$, the position of the migrating region for larger $K$, its progressive narrowing with increasing $K$, and its characteristic downward-sloping shape in the $N^S-N^L$ plane. These results demonstrate that the population-dependent escape rates provide a minimal description of the phase behavior underlying the granular clock dynamics.

\section{Discussion}
An important methodological aspect of the present study is the use of spatiotemporal diagrams to analyze granular-clock dynamics. To our knowledge, this type of representation has not been used previously in studies of granular-clock systems. Previous studies have instead typically presented the dynamics as time series of the particle number in each compartment. While such representations are useful for tracking population exchange, they provide only a limited view of the full spatiotemporal structure of the dynamics. By contrast, the spatiotemporal diagrams used here directly visualize the directionality and long-time persistence of cluster migration across multiple compartments. They also reveal weak gas-like particle emission as faint horizontal streaks, providing visual evidence that approximately symmetric particle emission persists even during sustained directional migration. These features are made particularly clear by the spatiotemporal representation used here.

On the modeling side, we constructed a flux-based model grounded entirely in experimentally measured escape rates, without explicitly incorporating detailed material properties or vibration parameters. Although characteristics such as particle diameter, mass density, restitution coefficient, and the external driving amplitude and frequency are not directly represented in the model equations, their influence is implicitly embedded in the escape rates of small and large particles. This formulation allows the model to remain simple while still capturing the essential dynamics of the system, as the escape rates effectively encode the combined influence of material and driving conditions on particle mobility. 

Population-dependent escape rates have also been examined in a previous numerical study of a granular clock system \cite{Li2012:_FluxMeasurement}. In that three-dimensional event-driven simulation, particle fluxes were measured to characterize the dependence of escape rates on the populations of small and large particles, and the qualitative trends are broadly consistent with those observed experimentally in Fig. \ref{fig:7}(b). However, their flux data were not incorporated into a dynamical model of cluster motion. In contrast, our measurements were conducted under the same physical conditions as the granular clock experiments, and the experimentally determined escape-rate functions were directly integrated into a dynamical model that successfully reproduced the observed directional cluster motion.

Flux-based models have frequently been used to describe particle transport in compartmentalized granular systems \cite{Eggers:_SandDemon,Lohse2004:_Coarsening, Lambiotte2005:_GranularClockSIM,Lohse2007:_Review}. For spontaneous cluster formation in monodisperse systems, Eggers derived the escape rate from a compartment using kinetic theory \cite{Eggers:_SandDemon, vdWeele:_Review}. In this framework, the number of particles escaping from a compartment with slit height $h$ per unit time in a two-dimensional system is given by
\begin{equation}
F(N)=F_0 N^2 \exp\left(-\frac{4 \pi g h r ^2 (1-e^2)}{v_b^2}N^2\right), \label{eq:eggers}
\end{equation}
where $N$ is the number of particles in the compartment, $F_0$ is a prefactor that sets the overall flux scale and depends on particle properties and driving conditions, $r$ is the particle radius, $e$ is the restitution coefficient, and $v_b=2 \pi f a$ is the peak velocity of the container bottom \cite{vdWeele:_Review,Eggers:_SandDemon}. This escape-rate framework has also served as the basis for flux models of granular-clock oscillations in compartmentalized systems.
When Lambiotte {\it et al.} extended it to bidisperse systems, they introduced additional variables to account for segregation effects in the escape dynamics \cite{Lambiotte2005:_GranularClockSIM,Viridi:_GranularClockEX,Lohse2007:_Review}. However, oscillatory behavior has also been observed in systems where particles differ only in mass density but not in size \cite{MHou2008:_Temperature}, indicating that segregation is not a necessary condition. Moreover, as shown in Fig. \ref{fig:1}(b), no clear vertical size segregation is present during translational cluster motion in our experiments. These observations suggest that segregation alone is not the fundamental mechanism underlying directional cluster migration. In contrast, our experimentally measured escape rates naturally encode this asymmetric interaction, allowing directional cluster motion to emerge without imposing additional phenomenological coupling assumptions.

It is also instructive to consider the possible physical interpretation of the fitting parameters in the escape-rate function. In Eggers’ kinetic-theory description, the escape rate exhibits the characteristic dependence $F(N) \propto N^2  \exp (-\alpha N^2)$, where comparison with Eq. (10) shows that $\alpha$ is proportional to the slit height $h$. In our formulation, the escape rate of small particles is described by the form 
$F(N) \propto N^2  [\exp (-BN) + \exp (DN)]$ (Eq. (1)).
The parameter $B$ may be associated with the actual slit height h, analogous to the role of $\alpha$ in Eggers’ description, whereas $D$ may reflect an additional population-dependent correction beyond the Eggers-type term. A positive value of $D$ may reflect an overflow effect at high $N$: as the number of particles increases, the height of the granular layer rises, which may alter the effective slit condition and lead to a rapid increase in the escape rate. Indeed, the fitted parameter $D^S (N^L)$ decreases monotonically with increasing $N^L$, as shown in Fig. \ref{fig:9}, indicating that this overflow-like contribution is itself strongly modulated by the surrounding particle population.
Such empirical relationships between fitting parameters and particle populations may provide useful guidance for developing a more physically grounded theory of escape rates in vibrated granular systems. While the first term is motivated by kinetic-theory considerations, a rigorous derivation of the full functional form remains an open problem. Further progress in connecting the empirical escape-rate function to kinetic-theory-like forms will likely require systematic experiments that vary accessible control parameters, such as the driving strength, slit height, particle diameter, and compartment base area.

While a more physically grounded derivation of the escape-rate function remains an important goal, it may lead to an expression different from the empirical one proposed here. Moreover, exact agreement between the experimental data and the fitted escape-rate functions is not necessary for capturing the observed cluster dynamics. Instead, the escape-rate functions need only preserve the essential population dependence. Crucially, they must retain the key asymmetry that large particles enhance the escape of small particles, whereas small particles suppress the escape of large particles. Under this condition, simplified functional forms may suffice. 
The measured escape-rate asymmetry is reminiscent of coupled activating and inhibitory interactions between particle species, analogous in spirit to those found in reaction-diffusion dynamics \cite{Turing:_ReacDiff,Meinhardt:_ReacDiffReview,Murray2003} and in population-dynamics models such as the Lotka-Volterra equations \cite{Goel1971}.
A key difference, however, is that the present system is a particle-number-conserving transport system: the dynamics arise from asymmetric inter-compartmental fluxes rather than from local production or annihilation terms. In this sense, directional cluster migration may be viewed as a conserved, transport-driven analogue of self-organized spatiotemporal dynamics more commonly discussed in non-conservative systems. A systematic stability analysis of the coupled differential equations could therefore provide further insight into the bifurcation leading to directional cluster migration.

A further point concerns the range of validity of the present fitting form, particularly in the low-$N$ regime. In this regime, the measured escape rate is better described by $F(N) \propto N^2  \exp (-BN)$, corresponding to the first term in our fitting function in Eq. (1). For reference, a representative fit comparison for the small-particle escape rate is provided in Supplementary Materials B. In the extremely dilute limit ($N  \ll 100$), the quadratic dependence predicted by the kinetic-theory expression is expected to hold. However, the flux becomes too small in this regime to be determined reliably in the present experiments. Because the cluster dynamics studied here involve relatively dense particle populations, the detailed functional form in the very dilute regime has little influence on the dynamics of interest. 

From a broader perspective, particle escape over the slit may be viewed as an activated process. In equilibrium statistical mechanics, the probability that a particle crosses an energy barrier is determined by the thermal temperature through a Boltzmann factor. In vibrated granular systems, however, the system is intrinsically out of equilibrium, and a well-defined thermodynamic temperature does not generally exist. Nevertheless, the concept of an effective granular temperature, often defined through velocity fluctuations, has been widely used to characterize particle agitation in driven granular media \cite{Goldhirsch2008}. A previous study of a bidisperse granular clock composed of equal-sized grains interpreted the oscillatory dynamics in terms of composition-dependent granular temperatures of the two species and their temporal oscillations \cite{MHou2008:_Temperature}. The present results suggest a complementary perspective in which the experimentally measured escape-rate functions provide an empirical bridge to an effective granular temperature, without requiring the granular temperatures to be quantified directly. If the escape probability over the slit can be related more explicitly to an activated-process framework, it may become possible to define an effective granular temperature directly from barrier-crossing dynamics rather than from velocity fluctuations. Notably, existing kinetic-theory-based escape-rate expressions such as Eq. (\ref{eq:eggers}) already contain explicit dependence on driving, dissipation, and particle parameters, suggesting that barrier-crossing formulations may offer a natural route toward an effective temperature that reflects both energy input and dissipation.

In previous studies, granular-clock oscillations were observed both when particles differed in size but were made of the same material \cite{Viridi:_GranularClockEX}, and when particles had the same size but different mass densities \cite{MHou2008:_Temperature}. Although we used silicon nitride and zirconia spheres to enhance visual distinction, we confirmed that directional cluster migration occurs in up to seven compartments even when both the small and large particles are made of zirconia (see Supplementary Materials C). These results indicate that the directional cluster migration is not specific to particular material combinations but rather reflects the collective escape dynamics governed by particle properties and interspecies interactions.

Environmental conditions, in particular ambient humidity, may influence the escape dynamics in vibrated granular systems. In the original experimental demonstration of the granular clock by Viridi et al. \cite{Viridi:_GranularClockEX}, electrostatic charging was mitigated by directing a faint stream of humid air onto the apparatus, although the relative humidity was not explicitly specified. In our experiments, measurements were conducted at room temperature under ambient humidity conditions typically ranging between $60 \%$ and $70 \%$. While strict humidity control was not implemented, care was taken to avoid excessively dry conditions that would enhance electrostatic effects.

Electrostatic charging can influence particle transport in two competing ways. Enhanced electrostatic repulsion between particles may increase their mobility and thereby enhance the escape flux. Conversely, adhesion of particles to the container walls can effectively reduce the number of mobile particles within a compartment, altering the apparent escape rate. Consequently, variations in humidity can lead to quantitative changes in the measured escape rates and, accordingly, in the characteristic migration time of the cluster. Indeed, we observed that the average time required for the cluster to migrate by one compartment exhibits moderate sensitivity to ambient conditions. Nevertheless, the qualitative structure of the escape-rate functions and the resulting phase diagrams remain robust, indicating that the directional cluster dynamics are governed primarily by the collective escape mechanism rather than by specific environmental details.

\section{Conclusion}
This study experimentally demonstrates sustained directional cluster migration in vibrated granular systems across up to seven compartments. By systematically varying the populations of small and large particles, we establish, for the first time, experimental phase diagrams in the $N^S-N^L$ plane for compartmentalized bidisperse granular systems. These measurements show that even in the well-studied case of $K=2$, the phase diagram contains two distinct migrating regions, a feature not previously identified experimentally. The direction of cluster migration remained unchanged even after a temporary interruption of vibration, providing direct evidence that the dynamics are governed by the instantaneous particle populations rather than by the prior history. By directly measuring the population-dependent escape rates, we constructed a minimal flux-based model from experiment that captures both the sustained directional migration and the overall phase-diagram structure. Taken together, these results show that extending the system to larger numbers of compartments was not merely a quantitative extension of earlier granular-clock studies, but a crucial step toward clarifying the population-dependent escape dynamics underlying directional cluster migration.

Our measurements reveal a clear asymmetric interaction between particle species: the escape of small particles is enhanced by the presence of large particles, whereas the escape of large particles is suppressed by small particles. This dynamical feedback naturally generates the directional cluster motion observed in the system. Because the escape rates are obtained directly from experiments, the resulting model implicitly incorporates the combined effects of particle properties, system geometry, and external driving conditions.

An important remaining challenge is to develop theoretical or empirical frameworks that explicitly relate these escape rates to particle properties such as mass, restitution coefficient, and friction, as well as to geometric parameters including slit height and cell depth. Establishing such connections would provide a more general predictive description of transport processes in vibrated granular systems.

\section*{Acknowledgements}
We are grateful to Hiroyuki Kitahata for generously lending us the electromagnetic shaker used in the experiments while our own device was out of order. We also thank Michio Otsuki and Jun-ichi Fukuda for their careful reading of the manuscript and valuable comments. We also thank Yasuyuki Kimura for valuable discussions and helpful advice throughout this study. S.I. and H.E. acknowledge support from the Japan Society for the Promotion of Science (JSPS) KAKENHI (Nos. 22K03468 and 22K03552) and the JSPS Core-to-Core Program, “Advanced core-to-core network for the physics of self-organizing active matter” (No. JPJSCCA20230002).

\bibliographystyle{apsrev4-2}
\bibliography{GranularClock}

@ARTICLE{Aranson:_PatternFormation,
  author = {I. S. Aranson and Lev S. Tsimring},
  title = {Patterns and collective behavior in granular media: Theoretical concepts},
  journal = {{R}ev. {M}od. {P}hys.},
  year = {2006},
  volume = {78},
  pages = {641}
}

@ARTICLE{Ottino:_SegregationReview,
  author = {J. M. Ottino and D. V. Khakhar},
  title = {Mixing and Segregation of Granular Materials},
  journal = {{A}nnu. {R}ev. {F}luid {M}ech.},
  year = {2000},
  volume = {32},
  pages = {55}
}

@ARTICLE{Goldhirsch:_GranularGas,
  author = {I. Goldhirsch and G. Zanetti},
  title = {Clustering instability in dissipative gases},
  journal = {{P}hys. {R}ev. {L}ett.},
  year = {1993},
  volume = {70},
  pages = {1619}
}

@article{Kudrolli1997,
  author  = {A. Kudrolli and M. Wolpert and J. P. Gollub},
  title   = {Cluster Formation due to Collisions in Granular Material},
  journal = {Phys. Rev. Lett.},
  volume  = {78},
  number  = {7},
  pages   = {1383--1386},
  year    = {1997},
  doi     = {10.1103/PhysRevLett.78.1383}
}

@ARTICLE{Eggers:_SandDemon,
  author = {J. Eggers},
  title = {Sand as Maxwell's Demon},
  journal = {{P}hys. {R}ev. {L}ett.},
  year = {1999},
  volume = {83},
  pages = {5322}
}

@ARTICLE{Lohse2002:_Collapse,
  author = {D. van der Meer and K. van der Weele and D. Lohse},
  title = {{S}udden collapse of a granular cluster},
  journal = {{P}hys. {R}ev. {L}ett.},
  year = {2002},
  volume = {88},
  pages = {174302}
}

@ARTICLE{Lohse2007:_Review,
  author = {D. van der Meer and K. van der Weele and P. Reimann and D. Lohse}, 
  title = {{C}ompartmentalized granular gases: flux model results},
  journal = {{J}.~{S}tat.\ {M}ech.},
  year = {2007},
  volume = {},
  pages = {P07021}
}

@ARTICLE{Lohse2004:_Coarsening,
  author = {D. van der Meer and K. van der Weele and  D. Lohse},
  title = {{C}oarsening dynamics in a vibrofluidized compartmentalized granular gas},
  journal = {{J}.~{S}tat.\ {M}ech.},
  year = {2004},
  volume = {},
  pages = {P04004}
}

@ARTICLE{Lohse2004:_Ratchet,
  author = {D. van der Meer and P. Reimann and  K. van der Weele and D. Lohse},
  title = {{S}pontaneous Ratchet Effect in a Granular Gas},
  journal = {{P}hys. {R}ev. {L}ett.},
  year = {2004},
  volume = {92},
  pages = {184301}
}

@ARTICLE{Lambiotte2005:_GranularClockSIM,
  author = {R. Lambiotte and J. M. Salazar and L. Brenig},
  title = {{F}rom particle segregation to the granular clock},
  journal = {{P}hys.\ {L}ett. A},
  year = {2005},
  volume = {343},
  pages = {224}
}

@ARTICLE{Viridi:_GranularClockEX,
  author = {S. Viridi and M. Schmick and M. Markus},
  title = {{E}xperimental observations of oscillations and segregation in a binary granular mixture},
  journal = {{P}hys. {R}ev. {E}},
  year = {2006},
  volume = {74},
  pages = {041301}
}

@ARTICLE{vdWeele:_Review,
  author = {K. van der Weele},
  title = {{G}ranular gas dynamics: How Maxwell's demon rules in a non-equilibrium system},
  journal = {{C}ontemp. {P}hys.},
  year = {2008},
  volume = {49},
  pages = {157}
}

@ARTICLE{Lohse:_MonoBifurcation,
  author = {D. van der Meer and K. van der Weele and D. Lohse},
  title = {Bifurcation diagram for compartmentalized granular gases},
  journal = {{P}hys. {R}ev. {E}},
  year = {2001},
  volume = {63},
  pages = {061304}
}

@ARTICLE{Turing:_ReacDiff,
  author = {A. M. Turing},
  title = {The chemical basis of morphogenesis},
  journal = {Philos. Trans. R. Soc. London, Ser. B},
  year = {1952},
  volume = {237},
  pages = {37}
}

@ARTICLE{Meinhardt:_ReacDiffReview,
  author = {A. Gierer and H. Meinhardt},
  title = {A theory of biological pattern formation},
  journal = {{K}ybernetik},
  year = {1972},
  volume = {12},
  pages = {30}
}

@book{Murray2003,
  author    = {J. D. Murray},
  title     = {Mathematical Biology II: Spatial Models and Biomedical Applications},
  edition   = {3rd},
  publisher = {Springer},
  address   = {New York},
  year      = {2003},
  series    = {Interdisciplinary Applied Mathematics},
  volume    = {18}
}

@article{Goel1971,
  author  = {N. S. Goel and S. C. Maitra and E. W. Montroll},
  title   = {On the Volterra and Other Nonlinear Models of Interacting Populations},
  journal = {Rev. Mod. Phys.},
  volume  = {43},
  number  = {2},
  pages   = {231--276},
  year    = {1971},
  doi     = {10.1103/RevModPhys.43.231}
}

@ARTICLE{MHou2008:_Temperature,
  author = {M. Hou and H. Tu and R. Liu and Y. Li and K. Lu and P.-Y. Lai and C. K. Chan},
  title = {{T}emperature oscillations in a compartmentalized bidisperse granular gas},
  journal = {{P}hys. {R}ev. {L}ett.},
  year = {2008},
  volume = {100},
  pages = {068001}
}

@ARTICLE{KCChen2008:_3Cells,
  author = {K.-C. Chen and C.-C. Li and C.-H. Lin and L.-M. Ju and C.-S. Yeh},
  title = {{E}xperimental study of Three-compartmentalized Granular Follower},
  journal = {{J}.~{P}hys.\ {S}oc. {J}pn},
  year = {2008},
  volume = {77},
  pages = {084403}
}

@ARTICLE{Liu2008:_ThreeCompartments,
  author = {Y. Liu and Q.-S. Mu and T.-D. Miao and J.-H. Liao},
  title = {{E}xperimental investigation about cyclic oscillations of a binary vibrofluidized granular mixture in $N$ connected compartments},
  journal = {{E}urophys.~{L}ett.},
  year = {2008},
  volume = {84},
  pages = {14004}
}

@article{vanderWeele2001hysteretic,
  author  = {van der Weele, K. and van der Meer, D. and Versluis, M. and Lohse, D.},
  title   = {Hysteretic clustering in granular gas},
  journal = {Europhysics Letters},
  volume  = {53},
  number  = {3},
  pages   = {328--334},
  year    = {2001}
}

@ARTICLE{Li2012:_FluxMeasurement,
  author = {Y. Li and R. Liu and M. Shinde and M. Hou},
  title = {{F}lux measurement in compartmentalized mono-disperse and bi-disperse granular gases},
  journal = {{G}ranular~{M}atter},
  year = {2012},
  volume = {14},
  pages = {137}
}

@article{Goldhirsch2008,
  author  = {Isaac Goldhirsch},
  title   = {Introduction to granular temperature},
  journal = {Powder Technology},
  volume  = {182},
  number  = {2},
  pages   = {130--136},
  year    = {2008},
  doi     = {10.1016/j.powtec.2007.07.022}
}

\end{document}